\documentclass[useAMS,usenatbib]{mn2e}
\usepackage[dvips]{graphicx,graphics,epsfig}
\newcommand{\Teff}{\ensuremath{T_{\rm eff}}}

\newcommand\z{{\it Z} }

\newcommand\2{{\footnotesize II}}
\newcommand\1{{\footnotesize I}}
\newcommand\lam{\ensuremath{\lambda}}
\newcommand\vsini{\ensuremath{v_{\rm e}\sin{i}}}
\newcommand\kms{\ensuremath{\mbox{km s}^{-1}}}
\newcommand\logg{\ensuremath{\log_{10}g}}

\newcommand{\pp}{\ensuremath{\phantom{-}}}
\newcommand{\ph}[1]{\ensuremath{\phantom{#1}}}

\newcommand{\eg}{{\em e.g.}}
\newcommand{\ie}{{\em i.e.}}

\newcommand{\sref}[1]{Section~\ref{#1}}

\title[A-type supergiants in the SMC] {Characteristics and
classification of A-type supergiants in the Small Magellanic Cloud}

\author[C.\ J.~Evans and I.\ D.~Howarth]
{Christopher J.\ Evans$^1$\thanks{email: cje@ing.iac.es} and
Ian D.\ Howarth$^2$\thanks{idh@star.ucl.ac.uk}\\
$^1$Isaac Newton Group of Telescopes, Apartado de Correos 321, 38700 Santa
Cruz de La Palma, Canary Islands, Spain\\
$^2$Department of Physics and Astronomy, University College London, Gower
Street, London WC1E~6BT,~UK.} 

\date{Dates to be inserted}

\begin{document}

\maketitle

\begin{abstract}
We address the relationship between spectral type and physical
properties for A-type supergiants in the SMC.  We first construct a
self-consistent classification scheme for A~supergiants, employing the
calcium $K$ to H$\epsilon$ line ratio as a temperature-sequence discriminant.
Following the precepts of the `MK process', the same morphological
criteria are applied to Galactic and SMC spectra with the
understanding there may not be a correspondence in physical properties
between spectral counterparts in different environments.  We then
discuss the temperature scale, concluding that A~supergiants in the
SMC are systematically cooler than their Galactic counterparts at the
same spectral type, by up to $\sim$10$\%$.  Considering the relative
line strengths of H$\gamma$ and the CH $G$-band we extend our study to
F and early G-type supergiants, for which similar effects are found.
We note the implications for analyses of extragalactic luminous supergiants, 
for the flux-weighted gravity-luminosity relationship and 
for population synthesis studies in unresolved stellar systems.
\end{abstract}

\begin{keywords}
Galaxies: individual: Magellanic Clouds --  
stars: fundamental parameters -- 
stars: early-type
\end{keywords}

\section{Introduction}
\label{intro}
A-type supergiants are important probes of metallicity in nearby
galaxies \citep[e.g.,][]{ml95, venn00b, bg02}.  This is because of
their intrinsic luminosity and relatively small bolometric corrections
(ensuring that they are among the visually brightest stars in any
galaxy; e.g., \citealt{hum83}), and their spatial isolation (which
contrasts with O-~and B-type stars, which are frequently found in
clusters or associations, or in binary systems with companions of
comparable brightness).  Moreover, not only are the optical spectra of
A-type supergiants typically free of significant contamination from
spatially unresolved companions, but they also exhibit lines from a
relatively wide range of elements, which can be modeled reasonably
satisfactorily using relatively simple LTE atmospheres
\citep[cf.][]{venn95}.

A-type stars can also be important -- occasionally dominant --
contributors to the integrated light of more distant, unresolved
galactic systems \citep[e.g.,][]{bb99}.  In the interpretation of such
unresolved systems, as well as the spectra of individual stars, the
correspondence between spectral morphology and fundamental stellar
parameters is a key issue.  We address this issue in the present
paper, motivated by the need to classify the A-type stars observed in
our forthcoming 2dF survey of the luminous-star content of the SMC \citep{ehi},
and to assign consistent and accurate effective temperatures to them.

\section{Spectral classification: principles}

\label{principles}

By way of introduction to the philosophy of spectral classification
which we adopt, we can do no better than draw on the discussion given
by \citet{m37} in his seminal paper on the classification of A--K-type
spectra:
\begin{quote}
``There would be a number of advantages in having a two-dimensional
empirical spectral classification; the operations of the determination
of actual values of stellar temperatures and luminosities could then
be separate from the problem of classifying spectra.  Suppose that it
had been the custom to publish the actual effective temperature
instead of the empirical spectral type{\ldots}the classification, then,
would be subject to two sources of error: (1) the error inherent in
the criteria themselves and the observational error introduced in
their estimation or measurement; and (2) the error introduced in the
reduction of the observational data to a temperature scale.  The
uncertainties introduced in (2) increase unnecessarily the uncertainty
in the actual operation of classification.''
\end{quote}
Thus, to quote a more recent doyen of classification,
\begin{quotation}
``One of the basic underlying principles of the MK approach to spectral
classification has been the preservation of its independence from
theoretical or other external information, with regard to both the
systemic formulation and its practical application'' 
\end{quotation}
\citep{wal79}.  We stress this independence here because it contrasts
with the practice of some analysts who infer `spectral types' from
model-atmosphere analyses \citep[e.g., in the context of SMC A-type
supergiants,][]{venn99}.

Spectral classification, according to the precepts of the `MK
Process', therefore follows observed spectroscopic characteristics,
not inferred physical properties.  A corollary of this is that two
stars with the same spectral types according to the traditional
two-parameter MK scheme may have different physical properties if they
differ in a third parameter, such as metallicity.   We investigate
the extent of these differences, insofar as they apply to the SMC,
in the latter part of this paper.

\section{Observations}
\label{observations}

If the MK philosophy is to be extended to galaxies beyond the Milky
Way, then the spectral morphology of an A0 star in the SMC, for
example, {\em should} resemble that of an A0 star in the Galaxy, 
in so far as the projection of a multi-dimensional parameter
space onto a two-dimensional classification scheme allows;
\citet{djl} provides an exemplary illustration of this principle in
his discussion of SMC B-supergiant spectra.  Thus, notwithstanding the
well-established differences in metallicity between SMC and Galactic
stars, it is necessary to use Galactic standards as a starting point.

\subsection{Galactic standards}

We found rather few suitable digital spectra of A-type classification
standards in the literature, and so resorted to obtaining our own; to
ensure continuity at the extremes of the A-star sequence, we also
observed a handful of late-B and early-F stars.  Details of the
targets are given in Table~\ref{atargets}.

The first set of spectra was obtained in 2000 July, during the course
of another programme, using the 2.5-m Isaac Newton Telescope (INT)
with the Intermediate Dispersion Spectrograph (IDS).  A 400B grating,
500-mm camera, and $1024{\times}1024{\times}24\mu{\mbox{m}}^2$
Tektronix detector (windowed to 300 pixels in the spatial direction)
gave wavelength coverage of approximately \lam3700--4900\AA\ at $R
\simeq 1900$ (as determined from arc-lamp exposures).

Further observations were obtained with the 4.2-m William Herschel
Telescope (WHT) in 2001 August, again during another programme, using
the Intermediate-dispersion Spectroscopic and Imaging System (ISIS)
with 600B and 1200B gratings (according to the requirements of the
principal programme) with a
$4096{\times}2048{\times}13.5\mu{\mbox{m}}^2$ EEV12 CCD (windowed to
700 pixels in the spatial direction and binned by a factor of two in
the dispersion direction).  The wavelength range from the 600B grating
was \lam3750--5400\AA\ at $R \simeq 2900$; coverage with the 1200B
grating was \lam3910--4780\AA\ at $R \simeq 5800$.  To provide a more
uniform dataset, the WHT A-star spectra were smoothed to $R \simeq
1600$, to match the 2dF data, for classification and measurement
purposes.

\begin{figure*}
\begin{center}
\epsfig{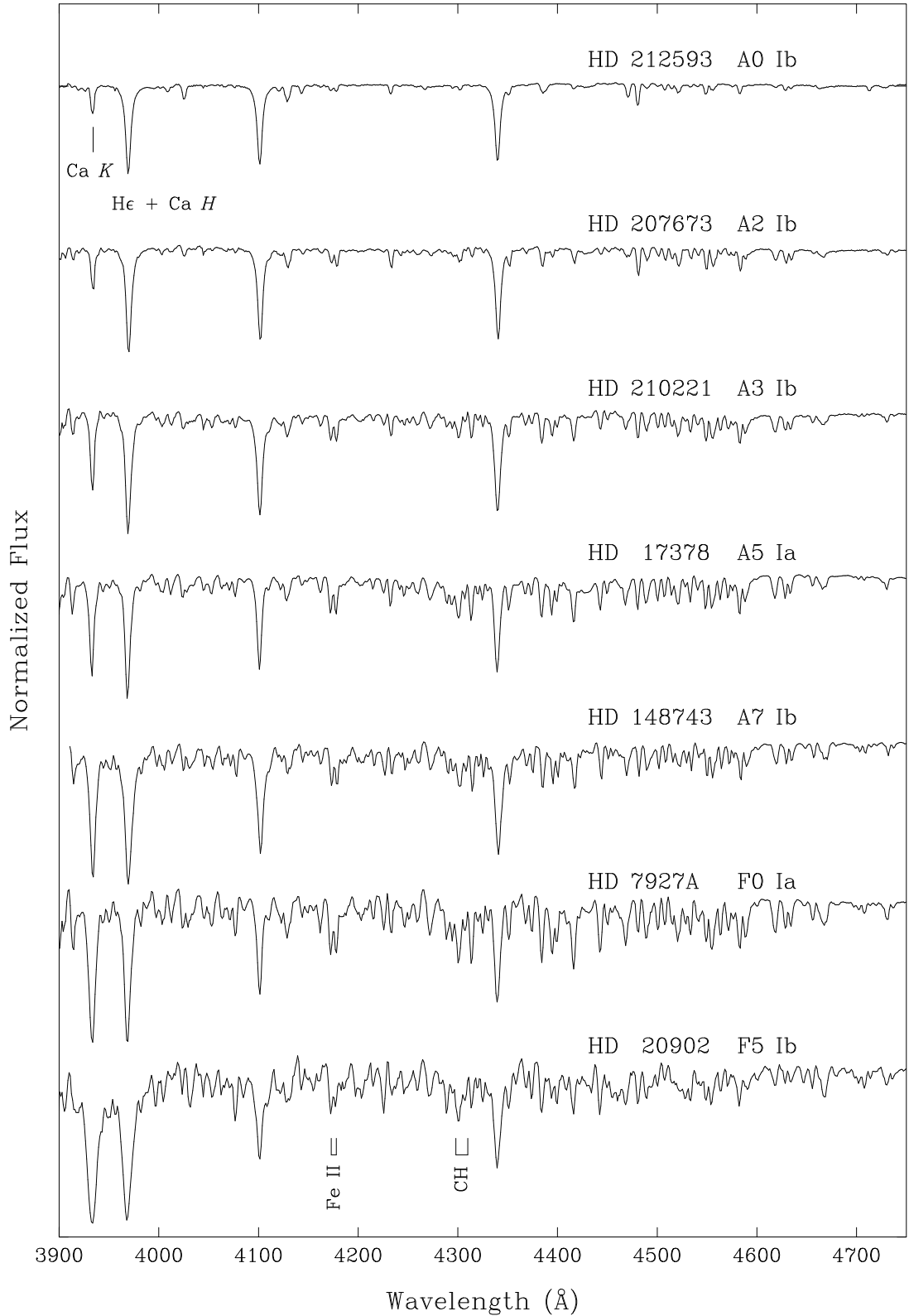}
\caption[]
{Spectra of Galactic A- and early-F-type stars.  Identified features
are Ca$\;$\2$\;K$ \lam3933; the Ca$\;$\2$\;$H \lam3969 + H$\epsilon$
blend; Fe$\;$\2 \lam4173-78; and the CH $G$-band, $\sim$\lam4300.
Successive spectra are offset by one continuum unit.}
\label{atypes}
\end{center}
\end{figure*}

The data typically have signal-to-noise ratios (s:n) $\ga$200
per wavelength bin.  Sample spectra are shown in Figure~\ref{atypes}.

\begin{table*}
\begin{minipage}{160mm}
\caption[]
{Galactic targets.  
Sources of spectral types are (in decreasing order of preference):
M73, \citep{mk73};
M43, \citep{mkk};
M55, \citep{mcw55};
M53, \citep{mhj53};
M50, \citep{mr50};
S54, \citep{slet54};
GG7, \citep{gg87};
GG9, \citep{gg89b};
G01, \citep{gnw01};
A81, \citep{abt81};
A85, \citep{abt85};
C69, \citep{ccjj}.
The photometric data in columns 6 \& 7 are taken from the
The General Catalogue of Photometric Data
\citep{mmh97},
and are used with intrinsic colours from \citet{fitz70}
to derive the estimated reddening values given in column~8, where we
retain unphysical negative values for statistical purposes.  
Where available, column 9 contains
\vsini~values (in \kms) from \citet{am95}. 
`Tel.' identifies the telescope, and, in the case of WHT observations, the grating
(600B or 1200B).
The final column is the ratio of the measured depth of the Ca$\;K$ line to that of the
Ca$\;H + {\rm H}\epsilon$ blend.

}
\label{atargets}
\begin{tabular}{lllcccccccccc}
\hline
%M50 112 362
%M55 2 41
$\ph{11}$HD  & $\quad$Name & Sp. Type & Source & $V$ & $(B-V)$ & $E(B-V)$ &
\vsini &Tel. & $\frac{1-{\rm I}_K}{1-{\rm I}_{(H+{\rm H}\epsilon)q}}$ \\
\hline
$\ph{111}$432  & $\beta$/11 Cas               &  F2$\;$III--IV &  M73         &  2.27 & 0.34       &$-0.03\pp$  & $\ph{1}$-- & INT & 1.01 \\
$\ph{111}$571  & 22 And                       &  F0$\;$II      &  M50         &  5.04 & 0.40       &  0.18      & $\ph{1}$-- & INT & 1.04 \\    
$\ph{11}$1404  & $\sigma$/25 And              &  A2$\;$V       &          GG7 &  4.52 & 0.06       &  0.01      & 110        & INT & 0.51 \\
$\ph{11}$3283  &                              &  A3$\;$II      &          GG9 &  5.79 & 0.28       &  0.21      & 100        & INT & 0.74 \\
$\ph{11}$3940  & V755 Cas                     &  A1$\;$Ia      &  M55         &  7.27 & 0.72       &  0.69      & $\ph{1}$-- & INT & 0.46 \\
$\ph{11}$6130  &                              &  F0$\;$II      &  M73         &  5.92 & 0.49       &  0.27      & $\ph{1}$23 & INT & 0.95 \\       
$\ph{11}$7927A & $\phi$/34 Cas                &  F0$\;$Ia      &  M55         &  4.99 & 0.68       &  0.53      & $\ph{1}$23 & INT & 1.00 \\
$\ph{11}$8538  & $\delta$/37 Cas              &  A5$\;$V       &  M53         &  2.68 & 0.13       &$-0.02\pp$  & 110        & INT & 0.77 \\
$\ph{1}$10845  & VY Psc                       &  A8$\;$III     &          GG9 &  6.58 & 0.25       &$-0.01\pp$  & $\ph{1}$85 & INT & 0.94 \\
$\ph{1}$12216  & 50 Cas                       &  A1$\;$V       &          S54 &  3.96 & $-0.01\pp$ &$-0.03\pp$  & $\ph{1}$85 & INT & 0.39 \\
$\ph{1}$12279  & 52 Cas                       &  A0$\;$V       &          GG7 &  5.99 & 0.03       &  0.04      & 255        & INT & 0.30 \\
$\ph{1}$12953  & V472 Per                     &  A1$\;$Ia      &  M55         &  5.69 & 0.62       &  0.59      & $\ph{1}$30 & W06 & 0.63 \\
$\ph{1}$13041  & 58 And                       &  A4$\;$V       &          GG9 &  4.82 & 0.12       &  0.00      & 120        & INT & 0.74 \\
$\ph{1}$13476  &                              &  A3$\;$Iab     &  M55         &  6.44 & 0.60       &  0.54      & $\ph{1}$20 & INT & 0.56 \\
$\ph{1}$14489  & $\iota$/9/V474 Per           &  A2$\;$Ia      &  M55         &  5.18 & 0.37       &  0.32      & $\ph{1}$25 & INT & 0.46 \\
$\ph{1}$15316  &                              &  A3$\;$Iab     &  M55         &  7.23 & 0.77       &  0.71      & $\ph{1}$-- & W06 & 0.67 \\
$\ph{1}$17378  & V480 Per                     &  A5$\;$Ia      &  M55         &  6.25 & 0.89       &  0.79      & $\ph{1}$25 & W06 & 0.82 \\
$\ph{1}$20902  &                              &  F5$\;$Ib      &  M73         &  1.79 & 0.48       &  0.22      & $\ph{1}$-- & INT & 1.02 \\
        148743 &                              &  A7$\;$Ib      &          G01 &  6.49 & 0.37       &  0.24      & $\ph{1}$43 & W12 & 0.95 \\
        161695 &                              &  A0$\;$Ib      &          C69 &  6.24 & 0.00       &  0.00      & $\ph{1}$25 & W12 & 0.27 \\
        172167 & $\alpha$/3 Lyr               &  A0$\;$V       &  M73         &  0.03 & 0.00       &  0.01      & $\ph{1}$15 & INT & 0.22 \\
        173880 & 111 Her                      &  A3$\;$V       &          GG9 &  4.36 & 0.13       &  0.05      & $\ph{1}$70 & W12 & 0.77 \\
        186155 &                              &  F5$\;$II      &  M73         &  5.07 & 0.39       &  0.01      & $\ph{1}$-- & INT & 1.02 \\
        187983 &                              &  A1$\;$Iab     &  M55         &  5.58 & 0.69       &  0.66      & $\ph{1}$45 & W12 & 0.49 \\
        192514 & 30 Cyg                       &  A3$\;$III var &          S54 &  4.82 & 0.10       &  0.01      & 160        & INT & 0.68 \\
        195324 & 42 Cyg                       &  A1$\;$Ib      &          C69 &  5.88 & 0.52       &  0.50      & $\ph{1}$15 & INT & 0.34 \\
        196379 &                              &  A9$\;$II      &  M73         &  6.13 & 0.40       &  0.22      & $\ph{1}$21 & INT & 0.95 \\
        197345 & $\alpha$/50 Cyg              &  A2$\;$Ia      &  M73         &  1.24 & 0.10       &  0.05      & $\ph{1}$30 & INT & 0.51 \\
        203280 & $\alpha$/5 Cep               &  A7$\;$IV--V   &  M53         &  2.46 & 0.22       &  0.00      & 180        & INT & 0.87 \\
        205835 & 74 Cyg                       &  A5$\;$V       &          GG9 &  5.04 & 0.17       &  0.02      & 185        & INT & 0.82 \\
        207260 & $\nu$/10 Cep                 &  A2$\;$Ia var  &  M55         &  4.29 & 0.52       &  0.47      & $\ph{1}$40 & INT & 0.46 \\
        207673 &                              &  A2$\;$Ib      &  M55         &  6.47 & 0.41       &  0.36      & $\ph{1}$35 & INT & 0.37 \\
        210221 & V399 Lac                     &  A3$\;$Ib      &  M55         &  6.14 & 0.42       &  0.36      & $\ph{1}$20 & INT & 0.65 \\
        211336 & $\epsilon$/23 Cep            &  F0$\;$V       &  M43         &  4.19 & 0.28       &$-0.04\pp$  & $\ph{1}$80 & INT & 0.96 \\
        212593 & 4 Lac                        &  B9$\;$Iab     &  M50         &  4.57 & 0.09       &  0.09      & $\ph{1}$-- & INT & 0.32 \\
        213558 & $\alpha$/7 Lac               &  A2$\;$V       &          S54 &  3.76 & 0.02       &$-0.03\pp$  & 115        & INT & 0.37 \\
        213973 &                              &  A9$\;$III     &          A81 &  6.01 & 0.32       &  0.04      & $\ph{1}$-- & W06 & 0.99 \\ 
        216701 & 1 Psc                        &  A7$\;$IV      &          GG9 &  6.11 & 0.20       &$-0.02\pp$  & $\ph{1}$80 & W06 & 0.93 \\
        222275 &                              &  A5$\;$III     &          A85 &  6.59 & 0.55       &  0.40      & $\ph{1}$-- & INT & 0.80 \\
        223385 & 6/V566 Cas                   &  A3$\;$Ia      &  M55         &  5.44 & 0.67       &  0.61      & $\ph{1}$30 & W06 & 0.76 \\
\hline
\end{tabular}

\smallskip
\textit{HD12953:} In contrast with the other Galactic standards, the
observed $K$/H$\epsilon$ ratio for this star is in relatively poor
accord with its MK spectral type (see Figure \ref{ca}); from the criteria
adopted here, HD 12953 would be classified as A3 -- with the exception
of slightly weaker metal lines and a smaller Ca $K$ intensity, its
spectrum is similar to that of HD 223385 (classified by M55 as A3 Ia).
However, 12953's A1$\;$Ia spectral type carries the authority of a
\citet{mcw55} classification.  The cause of this discrepancy is not
clear.

\smallskip
\textit{A note on `HD 187982':}
The HD catalogue assigns BD+24~3914 two numbers (HDs 187982 and
187983), and two spectral types (F5 and A2, respectively), noting that
`the spectrum is composite'.  \citet{hum78}, \citet{am95}, and others,
have subsequently identified the A-type star with HD~187982.  In her
study of spectroscopic binaries, \citet{h81}, also referring to
HD~187982, concluded that there is `no trace of a composite spectrum',
on the basis of greatly superior spectroscopic material, and she
mentions work by \citet{mcali79} in which the star was unresolved using
speckle interferometry with the Kitt Peak 4-m telescope.

It is important to note that the compilers of the HD catalogue
habitually assigned two numbers to what they considered to be single,
composite-spectra systems.  However, in this case it appears almost
certain that HD~187982 is non-existent, and that the A-type
supergiant BD+24~3914 should be identified with HD~187983 alone.

\end{minipage}
\end{table*}

\subsection{SMC data}
\label{smcdat}

As mentioned previously, the initial motivation for the present work was
the need to classify and interpret, in a systematic and consistent
manner, the A-star spectra obtained in our 2dF survey of the population
of luminous blue stars in the SMC.  The great bulk of our SMC dataset
therefore comes from this source, with observations obtained in 1998
September and 1999 September.

2dF is a fibre-fed, multi-object dual spectrograph mounted at the AAT
\citep{2df}.  We used the system with 1200B gratings, giving
wavelength coverage of $\sim$3900-4700 at $R \simeq 1600$.  The s:n
varies between $\sim$15--150 for the spectra retained for analysis,
averaging $\sim$25.
   
Because of initial concerns with the reliability of background
subtractions in the 2dF data, we obtained supplementary spectra of
$\sim$two dozen representative A-type SMC stars in traditional
long-slit mode in 2001 July, using the AAT's RGO spectrograph with a
1200B grating and EEV CCD ($R \simeq 2700$).  The observations have a
useful wavelength range of $\sim$3700--5500\AA, with s:n
$\sim$50--100.  They demonstrate that there are, in fact, no
systematic problems with the 2dF data.

\begin{figure*}
\begin{center}
\epsfig{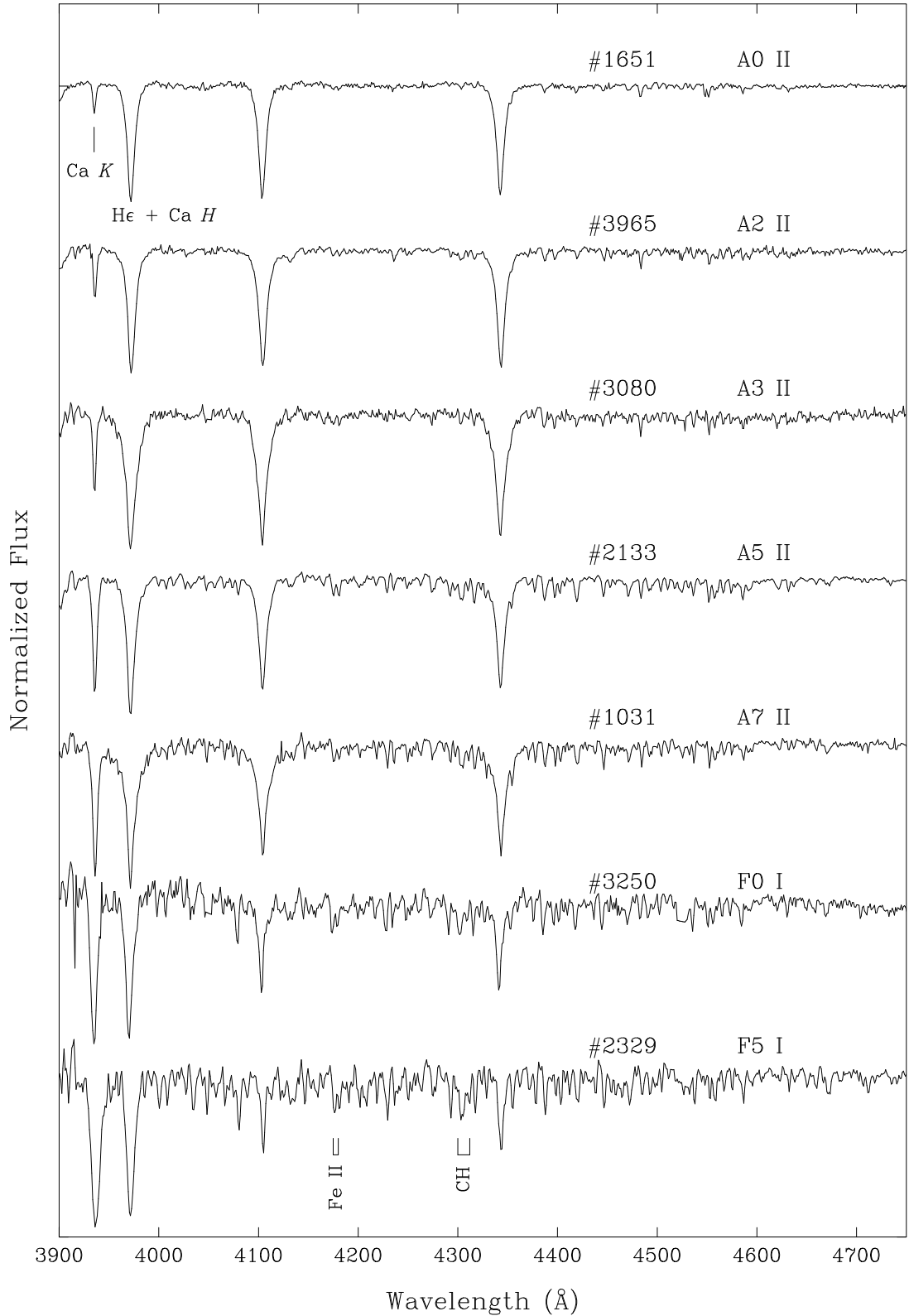}
\caption[]
{2dF spectra of representative A- and early F-type supergiants in the
SMC, identified by serial number in our forthcoming 2dF spectroscopic
catalogue.  Precise radial velocity measurements are not possible from
the 2dF data so the spectra are shown here have not been corrected to
the stellar rest frame.  The stars have been classified using the
criteria in Table~\ref{aclass}.}
\label{atypes2}
\end{center}
\end{figure*}

Illustrative 2dF SMC spectra are shown in Figure~\ref{atypes2}.

\section{Spectral classification: practice}
\label{classification}

\subsection{Luminosity classification} 
\label{lummore}
We begin our discussion by emphasising the distinction between
luminosity (intrinsic brightness) and luminosity {\em class}
(characterising the appearance of the spectrum).
Luminosity classes were allocated in our SMC work using the
\citet{azz87} calibration between luminosity class and H$\gamma$
equivalent width; mean values for the 
spectral types considered here are given in Table \ref{azzo_cal}.
Azzopardi's study was, necessarily, based on limited data
(low-dispersion, objective-prism photographic spectra acquired in the
1970s) and unfortunately the superior observational material of
\citet{bc74} does not extend beyond the earliest
A-subtypes.  We intend to revisit the relationships between H$\gamma$
line strength, luminosity class, absolute magnitude, and metallicity
in due course.  This is not necessary in the present,
morphologically-based, work, but attention is drawn to the discussion
in section~\ref{lumimp}.

The A-star spectra in Fig~\ref{atypes2} all have greater H$\gamma$
equivalent widths than those for the Azzopardi Ib calibration.  They
are also smaller than one would expect for class III spectra
\citep[\eg][]{bc74}, and have therefore been assigned class  II here.
In the absence of similar calibrations beyond A7, the two F-type
spectra are classified by extrapolation, as luminosity class I.

\begin{table}
\begin{center}
\caption{Mean H$\gamma$ equivalent width calibration for the MK `dagger'
spectral types used in the current work, from \citet{azz87}.}
\label{azzo_cal}
\begin{tabular}{cccc}
\hline
Sp. Type & Ia & Iab & Ib \\
\hline
A0 & 2.35 & 3.15 & 4.20 \\ 
A2 & 2.60 & 3.40 & 4.70 \\
A5 & 3.00 & 4.00 & 5.70 \\
A7 & 3.35 & 4.55 & 6.55 \\
\hline
\end{tabular}
\end{center}
\end{table}

\subsection{Temperature classification of A-type stars}

The foundations of modern spectral classification, elaborated by
\citet{mkk}, were largely in place by the time Annie Cannon produced
the Henry Draper Catalogue \citep[]{hdcat}, following earlier work by
Fleming, Maury and others \citep{pick90b, maury97}, in part under
Pickering's supervision.  In the A-type domain, Cannon's main
classification criterion was the intensity of the Ca$\;K$ line
(\lam3933\AA) with respect to H$\delta$ and the blend of H$\epsilon$ +
Ca$\;H$(\lam3968\AA).  \citet{m37} also identified the Ca$\;$\2 $K$
line as a useful diagnostic feature; it is the most conspicuous
discriminant of temperature for A-type stars.

We therefore investigated the use of the $K$ line and the
H$\epsilon$/Ca$\;H$ blend as a temperature discriminant. 
Although there
are, potentially, problems with the use of the $K$ line (discussed in
\sref{caprobs}), it has the advantage that a spectrum can be quickly
and accurately classified, even at low signal-to-noise.  While other
classification criteria exist, such as the ratio of Ca$\;$\1 \lam4226
to Mg$\;$\2 \lam4481 \citep{yam77}, the weakness of most metal-line
features in the spectra of (metal-deficient) SMC stars makes them less
useful for our purposes, especially in view of the modest s:n in the
majority of our 2dF spectra.

\subsubsection{Galactic stars}

The ratios of measured $K/{\rm H}\epsilon(+H)$ line depths in the
Galactic stars are given in Table~\ref{atargets}, and are shown in
Figure~\ref{ca} as a function of spectral type;\footnote{In principle,
the measured line-depth ratio may be influenced by spectral
resolution; in practice, we circumvent this problem by convolving all
our spectra to the resolution of the 2dF data.  The equivalent-width
ratio, which should be independent of resolution, proved to be a less
sensitive and consistent discriminant in practice.}  as expected,
there is a well-defined sequence.  However, there is some overlap
(e.g., between A1 and A2).  In view of this, and the relatively poor
s:n of the SMC data, we adopt a slightly less fine classification unit,
using only the `dagger' subtypes for which standards are advanced by
\citet{mk73}; the results are summarised in Table~\ref{aclass}, which
affords an almost completely `clean' segregation of the Galactic
standards (Fig.~\ref{ca}).

\subsubsection{SMC stars}

The criteria in Table~\ref{aclass} have been adopted for classifying
the SMC stars. Even for quite poor-quality data, the $K/{\rm
H}\epsilon$ ratio can be measured to better than $\sim$0.05, which is
generally sufficient for classifications uncertain to better than one
subtype at our adopted classification resolution.

The scheme in Table~\ref{aclass} is largely in accord with the
original criteria of Cannon \& Pickering.  
The sensitivity of $K/{\rm
H}\epsilon$ to spectral type falls off towards cooler temperatures,
and for both A9 and F0 types the $K$ line is very strong $(K/{\rm
H}\epsilon \simeq 1)$; these subtypes are essentially
indistinguishable in the 2dF data, and we grouped them under the `F0'
label.  (Still later spectral types are discussed in section~\ref{ftypes})

\begin{figure*}
\begin{center}
\epsfig{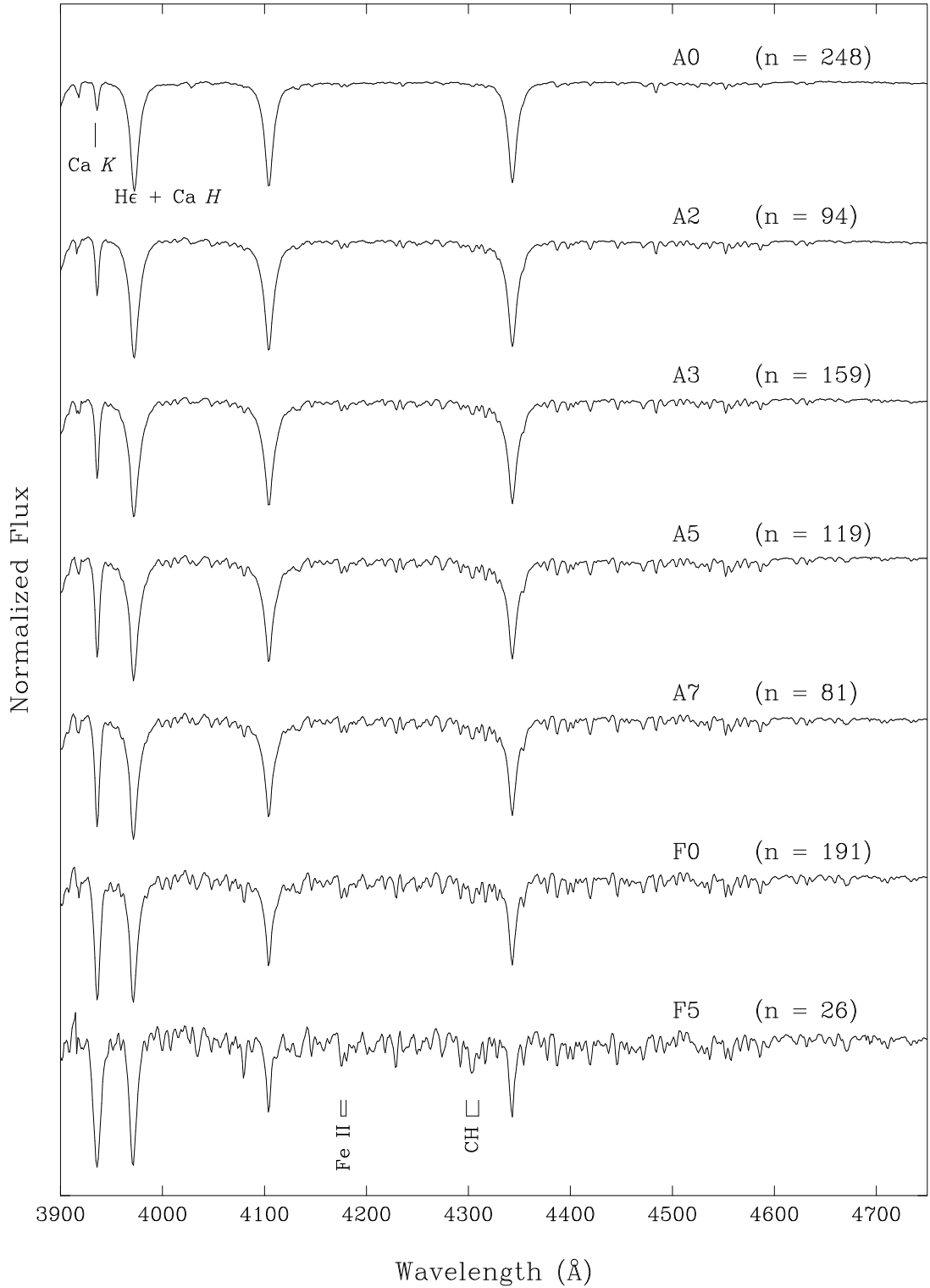}
\caption[] {Means of 2dF spectra for AF stars in the SMC.  
The numbers of individual spectra contributing to the means are given
in parentheses.}
\label{atypes3}
\end{center}
\end{figure*}

\begin{figure}
\begin{center}
\epsffile{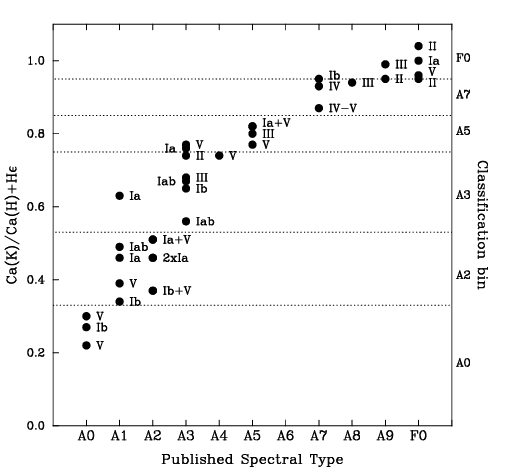}
\caption[]
{Ratio of core depth of Ca$\;K$ to Ca$\;H$ + H$\epsilon$, as a function
of spectral type, for the sample of Galactic A-type spectra listed in Table
\ref{atargets}.  Note that there is no obvious luminosity-class dependence of the ratio.
Dashed lines show the adopted classification boxes.   The most
discrepant point, at A1$\;$Ia, is HD~12953, discussed in Table~\ref{atargets}.}
\label{ca}

\epsffile{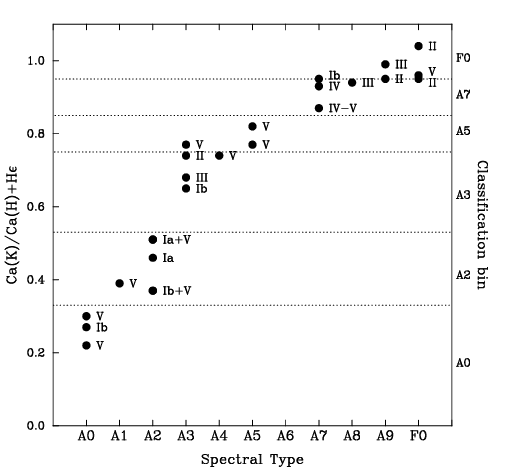}
\caption[]
{As in Figure~\ref{ca} except that only stars with values of $E(B-V)~<$~0.40 
are shown.}
\label{calowe}

\end{center}
\end{figure}

The distinction between A5 and A7 is relatively subtle, but is
achieved consistently.  This is illustrated by the signal-weighted
mean spectra at each spectral subtype shown in Fig.~\ref{atypes3}.
The smooth progression of weak metal features with spectral type
(e.g. Fe$\;$\2 \lam\lam4173-78) demonstrates that the classifications
are at least internally consistent, while discrimination at A3/A5/A7
is confirmed by the reversal of the $K/{\rm H}\delta$ ratio between
these subtypes.

\begin{table}
\begin{center}
\caption{Classification bins for A-type classifications, derived from 
observations of Galactic targets, using the system of \citet{mkk}.}
\label{aclass}
\begin{tabular}{c|l}
\hline
Ca$\;K$ / H$\epsilon$ + Ca$\;H$ & Spectral Type \\
\hline
$<$0.33 & A0 \\
0.33--0.53 & A2 \\
0.53--0.75 & A3 \\
0.75--0.85 & A5 \\
0.85--0.95 & A7 \\
$>$0.95 & A9, F0 or later \\
\hline
\end{tabular}
\end{center}
\end{table}
            
\subsection[]{Classification caveats}
\label{caprobs}
Although relatively quick and simple to use, the Ca~$K$ (and $H$)
lines are not without concerns for classification purposes.

\subsubsection{Luminosity effects}
\label{lumeff}
\citet{gg89b} noted that using the calcium/hydrogen ratio
may introduce effects into the spectral-type classifications because
the Balmer lines are sensitive to luminosity.  However, the
measurements in Table~\ref{atargets} show no obvious trend with
luminosity class, as is illustrated by in Fig.~\ref{ca}.  Although the
sampling is insufficiently dense to rule out small differences between
classes on average, the dispersion at a given spectral subtype is
evidently much larger than such differences.  In any case, our 2dF
sample is overwhelmingly dominated by high-luminosity objects.

\subsubsection{Interstellar contamination}
It has been known for a century that Ca$\;$\2\ $H$ \& $K$ absorption
lines arise in the interstellar medium \citep[][]{har04}.  The
interstellar lines are not distinguishable in intermediate-dispersion
spectra, and could therefore introduce a bias into the spectral
classifications.  (Although dust extinction towards SMC stars is
generally rather small, with $E(B-V) \la 0.1$, the velocity dispersion
in the line-of-sight interstellar material enhances the effect of
resonance-line absorption.)

To assess the likely impact of any contamination, reddening values
were calculated as a crude surrogate for interstellar-line equivalent
widths of Galactic targets (Table~\ref{atargets}).  Reassuringly,
there is no evidence of segregation between high- and low-reddening
targets (cf.\ Fig.~\ref{calowe}).   This is consistent with our
expectations that (i) interstellar absorption is a relatively small
contributor to the total $H, K$ equivalent widths for A-type stars,
and (ii) interstellar $H$ and $K$ will vary in tandem, so that they
have only a second-order effect on the $K/({\rm H}\epsilon + H)$
ratio.

\subsubsection{Metallicity effects}
\label{zeffects}
One of the most important considerations when using the calcium line
as a classification criterion is the reduced metallicity of the
SMC. One approach to the metallicity issue is given in \citet{venn99},
where the spectral types of a sample of SMC A-type supergiants were
adjusted in the light of model-atmosphere analysis.  Our approach
to the issues regarding classification that arise from this and other
studies were discussed in Section~\ref{principles}.  In the current
context of classifying the 2dF sample, the criteria in
Table~\ref{aclass} are applied on the basis that they are the {\it
morphological} counterparts of Galactic criteria, with the 
explicit understanding there may not be a correspondence in
{\it physical} properties between Galactic and SMC stars.

\subsubsection{Rotational broadening effects}
The influence of rotational broadening on A-stars classification has
been discussed by \citet{gg87}.  They defined low and high
\vsini~standards on the basis that broadening of the $K$ line in rapid
rotators may affect the classification process.  \citet{gg89b}
subsequently concluded that at type A5 and later this effect is
reduced, such that the Ca$\;K$/H$\epsilon$ ratio can be safely used
for rapid rotators.

This is not of direct concern in the 2dF dataset because of the
distance (i.e., luminosity) selection effect.  Typical rotational
velocities for A-type supergiants are low (of order 20--30 \kms; \eg,
\citealt{am95, venn95}), so rotation is not a significant issue.  For
comparison, the ``high'' \vsini~stars referred to by \citet{gg87} have
velocities in the range of 150--275 \kms (with luminosity types
ranging from III to V).

In principle there could be a systematic difference in the rotational
velocity distributions between the Galaxy and the SMC, as discussed
by \citet{mgm99} in reference to B-type stars.  Rotation could also
have an indirect effect through the Galactic stars used here to
describe the classification framework.  Where available, rotational
velocities from \citet{am95} are given in Table~\ref{atargets} for our
Galactic targets.  The majority of the Galactic stars have relatively
low velocities, and those with faster rotation are not particularly
discrepant in Figure~\ref{ca}, suggesting that rotational effects are
unlikely to play a significant r$\hat{\rm o}$le in the current work.

\subsection{Spectral classification of F- and later-type stars}
\label{ftypes}

The $K/{\rm H}\epsilon$ ratio is useful for distinguishing between A
spectral subtypes, even in data of moderate to poor quality, but is
not a sufficient criterion to distinguish late-B from early-A, 
nor  early-F from late-A
stars.

The B stars are straighforwardly identified by the presence of helium
lines \citep[e.g.][]{djl}, while the $\sim$100 supergiants with spectral types later than
F0 observed in the 2dF sample (as a result of photometric errors in
the input catalogue) were classified using the INT data as a point of
reference, in combination with the criteria given by \citet{jj90}.
Essentially, F-type stars are characterized by increasingly strong
metal lines (visible even at SMC metallicity); this behaviour is most
obvious in the ratio of Ca$\;K$ to H$\gamma$, but is clear in other
features, such as the \lam4300 $G$-band. The adopted `late-type'
classification scheme, which takes into account the characteristics of
the 2dF dataset (low s:n, weak metal lines), is summarised in
Table~\ref{fclass}.

\begin{table}
\begin{center}
\caption{Classification bins for F \& G-type classifications, derived
from \citet{jj90}.}
\label{fclass}
\begin{tabular}{c|l}
\hline
Spectral Type & Criteria \\
\hline
F0--F2 & Ca$\;K$ / H$\epsilon$ + Ca$\;H$ $\sim$1 \\
F5 & Clear presence of CH $G$-band \\
F8 & $G$-band / H$\gamma$ = 0.5--0.75  \\
G0 & $G$-band / H$\gamma$ = 0.75--0.9  \\
G2 & $G$-band $\sim$ H$\gamma$; H$\gamma$ $>$ Fe$\;$\1 \lam4325 \\
G5 &  H$\gamma$ $\sim$ Fe$\;$\1 \lam4325 \\
G8 &  H$\gamma$ $<$ Fe$\;$\1 \lam4325 \\
\hline
\end{tabular}
\end{center}
\end{table}

\section{The temperatures of A-type supergiants in the SMC}

For population syntheses, IMF modelling, and many other circumstances
where quantitative analyses of individual stars is infeasible, it is
important to have an accurate `look-up table' of correspondences
between physical and morphological characteristics.   In the remainder
of this paper, we examine this issue in the light of our remarks
in section~\ref{principles}, with particular emphasis on the
temperature scale for A-type stars.

\subsection{Model atmospheres}
\label{kurucz}
The model-atmosphere structures used here are  {\sc
Atlas}{\footnotesize 9} models \citep{k91}, calculated by
Collaborative Computational Project \#7 (CCP7;
{\tt http://ccp7.dur.ac.uk/}).  The models assume
plane-parallel geometry, hydrostatic equilibrium, and LTE, and include
the effects of line blanketing.  The majority of the A-type stars
in the 2dF sample are of luminosity types of Ib or II, where LTE has
been shown to be a reasonable approximation \citep[e.g.][]{pryz02}.  Departures
from LTE become increasingly important in type Ia supergiants;
however, in the current context, the {\sc Atlas}{\footnotesize 9}
models have the advantage that results can be achieved quickly,
relatively simply, and that they include the important effects of line
blanketing.

The low-metallicity CCP7 models that we used incorporate a
microturbulent velocity, $\xi$, of 2~\kms.  This is just below the
range of $\xi$~=~3--8 \kms~found in Venn's A-supergiant analyses
\citep[][1999]{venn95}, although many of her targets yielded values in
the 3--4 \kms~range.  Spectra synthesized using 2- and 4-\kms,
solar-abundance models are essentially identical, and differences
resulting from use of an 8-\kms~grid are negligible.  We conclude that
the CCP7 models are adequate for the current investigation.
 
Hubeny's unpublished {\sc synspec}{\footnotesize 43} code was used for
the spectral synthesis, running under the {\sc idl}
{\sc synplot} wrapper (also written by Hubeny).  
The necessary
atomic and molecular line-lists were downloaded from the
{\sc tlusty} web page.\footnote{{\tt http://tlusty.gsfc.nasa.gov/} at
the time of writing.}  Both Rayleigh-scattering and H$^-$-ion
opacities were included in the calculations.

The synthetic spectra were convolved with typical \vsini~values
($\sim{30}$~\kms), and smoothed to a resolution comparable to that of
the 2dF data.  In all cases, instrumental broadening dominates.

\subsection{Model grid}
Three main spectral regions were investigated:
\begin{itemize}
\item[--]{\lam3900--4050\AA~: Ca$\;K$/H$\epsilon$ (important in A-types).}
\item[--]{\lam4250--4450\AA~: CH $G$-band/H$\gamma$ (F and G-types).}
\item[--]{\lam4450--4540\AA~: He$\;$\1 \lam4471/Mg$\;$\2 \lam4481 (late-B stars).}
\end{itemize}
The atomic species included for the spectral-synthesis investigation
of B and A-types were H, He, C, N, O, Mg, Si and Ca.  For the later
types, in addition to the molecular line-list, Mn, Cr and Fe were included.

\begin{figure}
\begin{center}
\epsffile{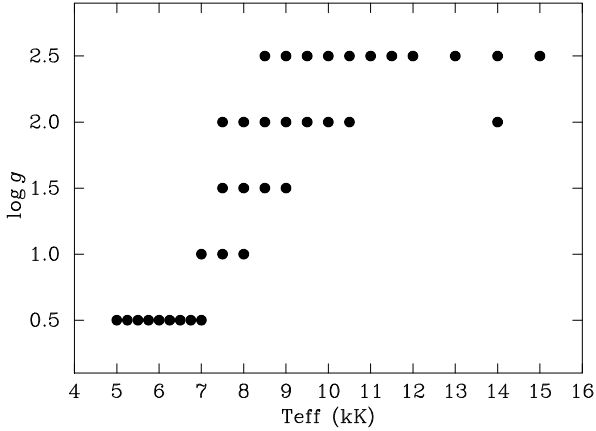}
\caption
{Grid of Kurucz models used to investigate the temperature scale of
BAF-type SMC supergiants.  The model atmospheres used were selected
with consideration to expected typical values, though the coverage of
the CCP7 grid is a limiting factor (\eg,~with the exception of \Teff~=
14kK,
the absence of \logg~=~2.0 at
higher temperatures).}
\label{grid}
\end{center}
\end{figure}

{\sc synplot} was used to generate synthetic spectra on a small grid
covering appropriate values of temperature and gravity.  The
temperature range of the grid was chosen by reference to 
\citet{sk82}.
Only approximate values are required for the \logg~sampling because
temperature has a more significant effect on the lines studied here.
Typical values of \logg~for A-type supergiants were estimated from
\citet{venn95} and \citet{vtg99b}.  \citet{duf00}
derive
\logg~$\approx~$2.3 for two B5 Iab stars, so \logg~=~2.5 was taken as
an upper limit for the hotter models.  For the F and G spectral types
considered here, \citet{lmb97} find values over a range of
\logg~=~0.0--1.0, so spectra were synthesized for our coolest models
using \logg~=~0.5.  The grid sampling is shown in Figure~\ref{grid}.

\subsection{Abundances}
\label{adopted}
For our calculations we have adopted a species-independent SMC
metallicity based on the median differential abundances found by
\citet{venn99} for SMC A~supergiants, i.e., $[X] = -0.76$
($\z_{\rm SMC} = 0.17\z_\odot$),
where
\[
[X] = \log_{10}\epsilon(X_{\rm SMC}) - \log_{10}\epsilon(X_{\rm Gal})
\]
for species $X$, and
\[
\log_{10}\epsilon(X) = \log_{10}\left({ n(X)/n({\rm H})}\right) + 12.
\]
The adopted metallicity is broadly consistent with other recently published values
for O and B-type stars (\citealt{hpl98}, \citealt{duf00}, \citealt{rvtd03}), 
F and K-type supergiants (\citealt{hill99}) and Cepheids (\citealt{lmb97}).

The metal abundances enter the modelling at two stages:
\begin{itemize}
\item[--]{In {\sc Atlas,} for the model-atmosphere calculation, $\z_{\rm Atlas}$;}
\item[--]{In {\sc synspec,} for the spectral-synthesis calculation, $\z_{\rm SSyn}$.} 
\end{itemize}
Although there are no CCP7 {\sc Atlas} models specific to $[X] =
-0.76$, grids at $[X] = -0.5$ and $-1.0$ (\ie~$\z_{\rm
Atlas}~=~0.32\z_{\odot}$ and $0.10\z_{\odot}$) span our adopted
metallicity.  Since the dominant effect of metallicity in the computed
spectra arises through
the spectrum
synthesis, not the atmospheric
structures, we can use these {\sc Atlas} models for our purposes.  
We adopted the 0.32$\z_{\odot}$ structure models, with 
0.17$\z_{\odot}$ for the spectral synthesis.

[The insensitivity of computed spectra to model structure is
illustrated in Figure~\ref{abund}, which shows model profiles for
$\z_{\rm Atlas} = \z_{\odot}, 0.32\z_{\odot}$, and $0.10\z_{\odot}$ at
\Teff = 9000K, \logg = 1.5, $\z_{\rm SSyn} = 0.10\z_{\odot}$.
When degraded to 2dF resolution, the resulting spectra are practically
indistinguishable; the model-atmosphere structures calculated with
different metallicities lead to essentially the same result, provided
that the abundance is held constant in the spectrum synthesis.
Similar results are obtained at \Teff~=~6000K, \logg~=~0.5, as shown
in Figure~\ref{abund3}, where the largest differences in rectified
intensity are of order 2--3$\%$ -- less than the uncertainties in the
observed 2dF spectra.

These results are underpinned by inspection of the temperature profiles
of the structure models.  The profiles of the 9000K models, shown in
Figure~\ref{atmos}, are nearly identical for all three metallicities.
The same is true of the 6000K models, shown in Figure~\ref{atmos2}.
Only at large depths is the solar-abundance model significantly offset, but, as
shown in Figure~\ref{abund3}, the effect is small in terms of the
emergent spectra.]

\begin{figure}
\begin{center}
\epsffile{abund.ps}
\caption
{Model spectra for the Ca$\;K$ region (\Teff~=~9000K, \logg~=~1.5),
degraded to the resolution of the 2dF data.  The metal abundances used
in the spectral synthesis are 0.10$\z_{\odot}$ in all cases; the
abundances in the model atmospheres are solar, 0.32$\z_{\odot}$, and
0.10$\z_{\odot}$.  The resulting spectra are essentially identical.}
\label{abund}

\epsffile{abund3.ps}
\caption
{Model spectra of the H$\gamma$ region (\Teff~=~6000K, \logg~=~0.5),
degraded to the resolution of the 2dF data; abundances are as in
Figure~\ref{abund}.  The resulting spectra are again very similar.}
\label{abund3}
\end{center}
\end{figure}

\begin{figure}
\begin{center}
\epsffile{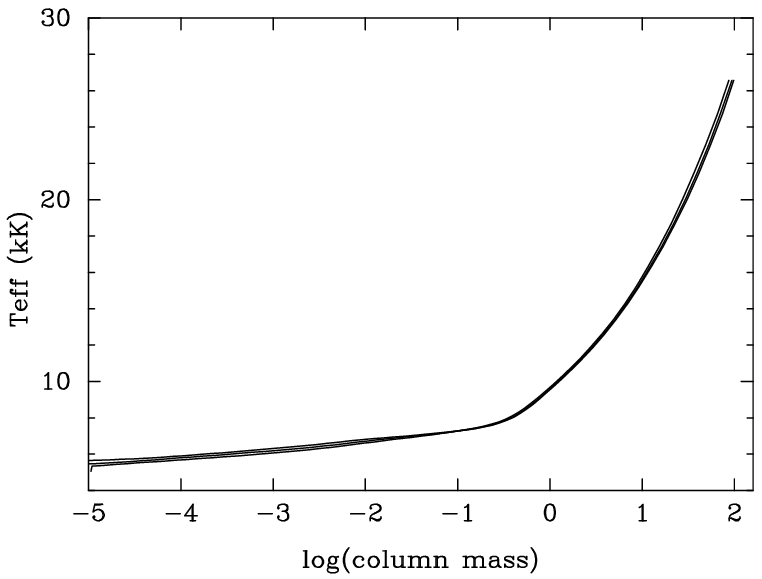}
\caption
{Atmospheric temperature profiles at \Teff~=~9000K, \logg~=~1.5.
Models shown have $\z_{\rm Atlas}$~=~$\z_{\odot}$, 0.32$\z_{\odot}$ and
0.10$\z_{\odot}$,  and are essentially identical.}
\label{atmos}

\epsffile{atmos2.ps}
\caption
{Atmospheric temperature profiles at \Teff~=~6000K, \logg~=~0.5.
Models shown have $\z_{\rm Atlas}$~=~$\z_{\odot}$, 0.32$\z_{\odot}$ and
0.10$\z_{\odot}$.  The models differ only at large depths (see
Section~\ref{adopted} for details).}
\label{atmos2}
\end{center}
\end{figure}

\subsection{Spectral-type--temperature calibration}

After rectification and smoothing, the synthetic spectra were assigned
`high-resolution' spectral types, using the criteria described in
\sref{classification}.  Temperatures were then interpolated to the
coarser grid of spectral types in Table~\ref{aclass}.  Galactic
(solar) results were obtained by using the same procedures, to provide
a directly comparable reference, and as a check on our methods.  The
calibrations are given in Table~\ref{scheme}, and directly
demonstrate the effect of SMC metallicity: at the same
morphological spectral type, SMC stars are systematically cooler
than Galactic counterparts, by up to $\sim$10{\%} at late~A.  

\begin{figure}
\begin{center}
\epsffile{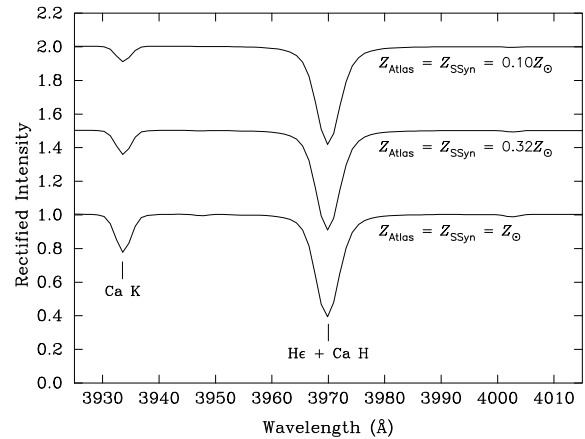}
\caption[]
{Model spectra for the Ca$\;K$ region (\Teff~=~9000K, \logg~=~1.5),
degraded to the resolution of the 2dF data.  The metal abundances used
in the spectral synthesis are identical to those used in the
model-atmosphere calculations and are as shown.}
\label{abund2}
\end{center}
\end{figure}

The explanation for this temperature offset is straightforward.  The
primary classification criterion used here involve the calcium and
Balmer spectral lines.  For sub-solar metallicity, the effective
temperature has to be lower in order to increase the Ca~$K$ line
strength to that required to satisfy the $K$/H$\epsilon$ criterion
established from Galactic standards.  This point is illustrated in
Figure \ref{abund2}, using models at constant temperature but
decreasing $Z$.  (It may be noted that decreased metallicity also
reduces the Balmer-line strength, because the pressure is reduced; but
this is a much smaller effect than the direct impact on metal-line
strengths.)

Given this interpretation, one might question whether the cooler SMC
A-star temperatures highlighted here are simply a consequence of the
particular classification criterion adopted.  However, our temperature
calibrations are in excellent agreement with both the detailed
analyses of individual Galactic stars given by \citet{venn95}, and
with her results for early-A SMC supergiants \citep{venn99}, which are
based largely on the Mg$^0$/Mg$^+$ ionization balance.  Nonetheless,
there are differences in philosophy; for example, \citet{venn99}
obtained \Teff~=~7900K (and \logg~=~0.9) for AV~442, and on that basis
reclassified the star from A3 to A7, implicitly adopting a
metallicity-independent relationship between spectral type and
temperature.  Our approach is to adopt spectral types consistently
based solely on spectral morphology, but to accept a
metallicity-dependent relationship between spectral type and physical
parameters.  That is, we would retain the A3 classification for
AV~442, but would assign $T_{\rm eff} =$ 8000~K (Table~\ref{scheme}).
Three stars in Venn's SMC sample appear to have temperatures that are
inconsistent with those in Table \ref{scheme} given their spectral
types, namely AV~136 (Sk 54, A0 Ia), 478 (Sk 154, A0 Ib) and 213 (Sk
75, A2 Iab).  These discrepancies appear to largely be a result of
different published classifications, e.g. Venn's temperatures are
consistent with the spectral types of \citet{hkg91} who gave
classifications for AV~478 (A5 Iab) and AV~213 (A2--3 I).

\begin{table}
\begin{center}
\caption{Solar-abundance and SMC temperature calibrations from the current work.}
\label{scheme}
\begin{tabular}{lll}
\hline
Type & Solar & SMC \\
\hline
B8 & 13000 & 12000 \\
B9 & 10750 & 10500 \\
A0 & {\ph 1}9750 & {\ph 1}9500 \\
A2 & {\ph 1}9000 & {\ph 1}8500 \\
A3 & {\ph 1}8500 & {\ph 1}8000 \\
A5 & {\ph 1}8250 & {\ph 1}7750 \\
A7 & {\ph 1}8000 & {\ph 1}7250 \\
F0 & {\ph 1}7500 & {\ph 1}6750 \\
F5 & {\ph 1}6250 & {\ph 1}5875 \\
F8 & {\ph 1}6000 & {\ph 1}5750 \\
G0 & {\ph 1}5875 & {\ph 1}5625 \\
G2 & {\ph 1}5750 & {\ph 1}5500 \\
G5 & {\ph 1}5500 & {\ph 1}5250 \\
\hline
\end{tabular}
\end{center}
\end{table}

\begin{figure}
\begin{center}
\epsffile{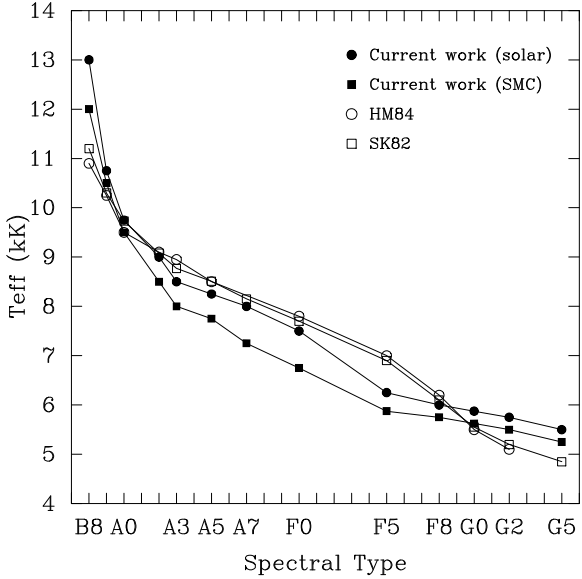}
\caption
{Solar and SMC temperature calibrations compared with those 
of Humphreys \& McElroy (1984) and Schmidt-Kaler (1982).  
The differences for later spectral types between the published 
calibrations and our solar values, notably at F5, are likely
attributable to difficulties of classification in this region and
to refinements in the model atmospheres codes used here.}
\label{scales}
\end{center}
\end{figure}

Figure~\ref{scales} compares the solar-abundance temperature scale
derived here with those compiled by \citet{sk82} and by \citet{hm84}.
The main differences are at the very earliest and latest types
considered here.  The higher values we find for the late B-types are
consistent with newer calibrations \citep[e.g.,][]{pc98}, so this is
not a cause for concern.  The differences in the later types, notably
F5, could be attributable to the difficulties of classification in
this region or to refinements in the model atmosphere codes (for
example, the inclusion of molecular lines).

For simplicity we assumed a single relative abundance for all metallic
species (i.e. 0.17$Z_\odot$); however, detailed analyses such as those
of \citet{venn99} find a spread of abundances for different species.
Venn specifically highlighted the relative underabundance of the alpha
elements, including calcium (in contrast to the SMC H~\2 region
results from \citet{r92} which found no calcium underabundance).  To
investigate the significance of adopting a lower calcium abundance, 
model spectra ($Z$ = 0.10$Z_\odot$) were calculated for 7500,
8000 and 8500~K, the domain in which the $K$ line is most sensitive to
temperature and abundance effects.  Such a reduction enhances the 
temperature effect described here, but the change is very small
compared to that between Galactic and `SMC' abundances.

\subsection{Luminosity implications}
\label{lumimp}

\begin{figure}
\begin{center}
\epsffile{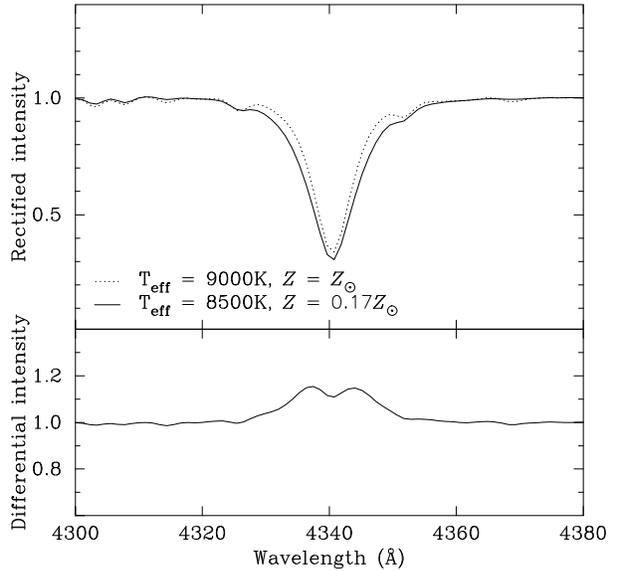}
\caption
{Model H$\gamma$ spectra at spectral type A2, for temperatures and
metallicities appropriate for Galactic and SMC supergiants
(with \logg~$=$~2.0).  The differential intensity plot in the lower panel
emphasises of the difference in line strengths.}
\label{ew}
\end{center}
\end{figure}

The present discussion of the relationship between temperature and
spectral type in the reduced-metallicity environment of the SMC has
immediate implications for the criteria used to allocate luminosity
classes.  At given spectral type, the temperature of an A-type SMC
star is lower than its Galactic counterpart; as a consequence, the
H$\gamma$ equivalent width will be different in the SMC star.
For example, Figure~\ref{ew} shows
the model H$\gamma$ line for \ie~\Teff~=~9000K, \z~=~\z$_{\odot}$ and
\Teff~=~8500K, \z~=~0.17\z$_{\odot}$ (at \logg~=~2.0), corresponding
to Galactic and SMC A2 spectra, respectively; the 
H$\gamma$ equivalent width is
some 30$\%$ larger in the SMC case.
This is not (directly) a metallicity effect, as the
H$\gamma$ lines for Galactic and SMC metallicities are essentially
identical at fixed temperature.

Table~\ref{factor} summarizes the differences in H$\gamma$ equivalent
widths for stars of spectral types B8--A7 (averaged over \logg~at a
given temperature).  In general, equivalent widths are
$\sim$10--30$\%$ larger in the synthetic SMC spectra than in Galactic
models at the same spectral type (but {\em not} the same temperature).
Note that the equivalent widths are larger in $all$ of the SMC spectra; 
in contrast to A-type dwarfs where the equivalent width of the Balmer
lines reaches a maximum around A0/A2, the maximum for supergiants 
occurs at a much later type \citep[e.g.][]{h66} resulting in a positive
effect throughout the spectral types in Table \ref{ew}.
The clear implication is that if luminosity classifications are
allocated from Galactic calibrations, the intrinsic brightness of a
star with a given classification cannot be assumed to be universal.

\begin{table}
\begin{center}
\caption
{Ratio of H$\gamma$ equivalent widths for Galactic and SMC model
spectra corresponding to the same morphological types.  The
differences arise because of the lower temperatures of the SMC models
at a given type, which, in the B8-A7 range results in increased
equivalent widths.}
\label{factor}
\begin{tabular}{cc}
\hline
Type & H$\gamma_{\rm SMC}$ / H$\gamma_{\rm Gal.}$ \\
\hline
B8 & 1.19\\
B9 & 1.08\\
A0 & 1.10\\
A2 & 1.29\\
A3 & 1.38\\
A5 & 1.27\\
A7 & 1.61\\
\hline
\multicolumn{2}{c}{} \\
\multicolumn{2}{c}{} \\
\end{tabular}
\end{center}
\end{table}

\section{Discussion}
\label{discuss}

Our work has demonstrated that accurate A-supergiant temperatures
require knowledge of the metallicity of a system (and vice versa).
This effect has important ramifications for studies of extragalactic
A-type supergiants.  For example, from comparison with template
Galactic spectra, \citet{bk01} assign a spectral type of A1 Ia to a
star in NGC 3621 (a spiral galaxy at a distance of 6.7 Mpc).  From the
spectral type they estimate a temperature (9000K $\pm$400) and then
use Kurucz model atmospheres and the line formation calculations of
\citet{pryz02} to find the chemical composition which best matches the
observations; their final model has a chemical composition comparable
to that of the Large Magellanic Cloud.  The source of their
temperature estimate is not given and unfortunately their spectra do
not extend blueward to the Ca~$K$ line (their Figure~2).  $If$
the temperature were from Galactic analyses then the lower metallicity
model would fail to reproduce the observed $K$-line intensity,
requiring a reduction in the model temperature, leading to a 
different abundance determination; i.e. some degree of iteration is
required.

Similar complications arise for the study of NGC 300 by
\citet{bg02}.  Detailed atmospheric parameters were given
for two supergiants, adopting temperatures from \citet{hm84} on the
basis of their spectral types.  The metallicity of their star `A-8' (B9--A0
Ia) star is equivalent to that of the SMC, with $[X] = -0.7 \pm0.2$.
In light of the temperature effect described here, the uncertainty in such 
abundance determinations will be larger than $\pm$0.2 dex; even moderate 
changes of temperature can have a significant impact on the derived 
abundances \citep[e.g.][ Table 3]{venn99}.

\begin{figure}
\begin{center}
\epsffile{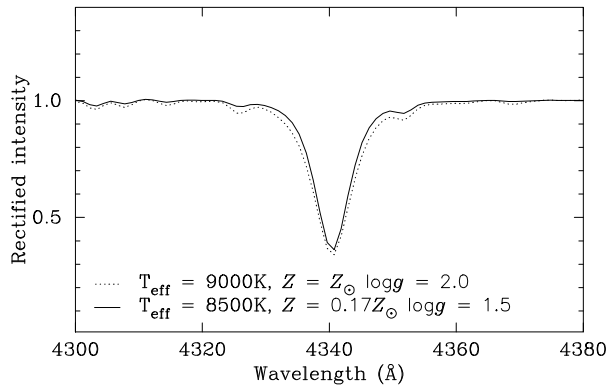}
\caption
{Model H$\gamma$ spectra at spectral type A2, for temperatures and
metallicities appropriate for Galactic and SMC supergiants but with
different gravities, \logg~$=$~2.0 and 1.5 respectively.  }
\label{hgamma2}
\end{center}
\end{figure}

The results presented here also have implications for the
`flux-weighted gravity-luminosity relationship' (FGLR) proposed by
\citet{k03}.  They argue that for a given effective temperature
(in their illustrative cases, assumed from the spectral type) the higher-order Balmer lines
(H$\gamma$ through to H11) can be used to determine the
gravity of distant A-type supergiants accurately.  After calibration, the FGLR
can in principle be used to determine the bolometric magnitude of a
given star, hence the distance.  Such a relationship is an attractive
proposition for extragalactic distance determinations, especially
since the atmospheric modelling only relies on the hydrogen lines
which are thought to be well understood.  However, in the context of
the current work, consider the case of an A2 type supergiant, for
which \citet{k03} adopt \Teff~$=~9000$K.  Such a temperature agrees
with our solar-abundance temperature in Table \ref{scheme}; however, at
SMC metallicity \Teff~$=~8500$K would be more appropriate.  Herein
lies a complication for the FGLR.  The H$\gamma$ profile for a
solar-abundance model with \Teff~$=~9000$K and \logg~$=$~2.0 is
similar to that for an SMC-abundance \Teff~$=~8500$K,
\logg~$=$~1.5 model (Figure \ref{hgamma2}); the difference in $M_{\rm bol}$ 
from the FGLR from these two models is 1.7$^{\rm m}$.  With a finer
sampling of gravity in the model grid the uncertainty in \logg~need
not be as extreme as 0.5; nonetheless, the estimate of 0.05 by Kudritzki et
al. must be considered rather optimistic.  Disentangling the effects
of temperature and metallicity in A-supergiants is non-trivial,
impacting on their potential for use as extragalactic distance indicators.

\section{conclusion}
\label{end}
We have observed a large number of A (and some later-type) supergiants
as part of our 2dF survey of the SMC.  To facilitate an accurate
application of the MK process to our digital data, Galactic standards
have been observed to provide a simple but quantitative
temperature-classification scheme; the 2dF spectra have been
classified within that scheme.  A direct result of classification
based purely on morphological criteria is that, because the primary
criterion involves both metallic and Balmer lines, the temperature of
an SMC A-type supergiant is $\sim$5--10$\%$ cooler than its Galactic
counterpart.  One of the main implications of this effect is that
Galactic temperature calibrations should not necessarily be used when
determining fundamental parameters and abundances for extragalactic A-supergiants.

In addition to its significance for individual stars, this result has
implications for population-synthesis studies.  Such investigations
typically involve generating a synthetic population from
stellar-evolution models, and then transforming the model predictions
to observational parameter space.  This transformation necessarily
always involves a calibration between observed and modelled
parameters; if the calibration is based on Galactic stars, then
significant errors can be introduced into the interpretation of
low-metallicity environments, even when the synthetic population is
generated using stellar-evolution models at the appropriate
metallicity.

\section{acknowledgements}
CJE was funded by PPARC during the course of this work.  IDH is a
Fresia Jolligoode Fellow, and acknowledges support from SWIMBO.  We
thank Ivan Hubeny and the CCP7 project for assistance with the
model-atmosphere work, and the staffs of the Anglo-Australian
Observatory and of the Isaac Newton Group for support at the
telescopes.  We are grateful to Kim Venn and Nolan Walborn for their
detailed comments on the manuscript.  This paper is based in part on
data obtained with the Isaac Newton and William Herschel Telescopes,
operated on the island of La Palma by the Isaac Newton Group in the
Spanish Observatorio del Roque de los Muchachos of the Instituto de
Astrofisica de Canarias, and with the Anglo-Australian Telescope.

\vspace{-0.25cm}
\bibliography{ms}

\begin{thebibliography}{}

\bibitem[\protect\citeauthoryear{{Abt}}{{Abt}}{1981}]{abt81}
{Abt} H.~A.,  1981, \apjs, 45, 437

\bibitem[\protect\citeauthoryear{{Abt}}{{Abt}}{1985}]{abt85}
{Abt} H.~A.,  1985, \apjs, 59, 95

\bibitem[\protect\citeauthoryear{{Abt} \& {Morrell}}{{Abt} \&
  {Morrell}}{1995}]{am95}
{Abt} H.~A.,  {Morrell} N.~I.,  1995, \apjs, 99, 135

\bibitem[\protect\citeauthoryear{{Azzopardi}}{{Azzopardi}}{1987}]{azz87}
{Azzopardi} M.,  1987, \aas, 69, 421

\bibitem[\protect\citeauthoryear{{Balona} \& {Crampton}}{{Balona} \&
  {Crampton}}{1974}]{bc74}
{Balona} L.,  {Crampton} D.,  1974, \mnras, 166, 203

\bibitem[\protect\citeauthoryear{{Bresolin}, {Gieren}, {Kudritzki},
  {Pietrzy\'{n}ski} \& {Przybilla}}{{Bresolin} et~al.}{2002}]{bg02}
{Bresolin} F.,  {Gieren} W.,  {Kudritzki} R.-P.,  {Pietrzy\'{n}ski} G.,
  {Przybilla} N.,  2002, \apj, 567, 277

\bibitem[\protect\citeauthoryear{{Bresolin}, {Kudritzki}, {Mendez} \&
  {Przybilla}}{{Bresolin} et~al.}{2001}]{bk01}
{Bresolin} F.,  {Kudritzki} R.-P.,  {Mendez} R.~H.,    {Przybilla} N.,  2001,
  \apj, 548, 159L

\bibitem[\protect\citeauthoryear{{Brotherton}, {van Breugel}, {Stanford},
  {Smith}, {Boyle}, {Miller}, {Shanks}, {Croom} \& {Filippenko}}{{Brotherton}
  et~al.}{1999}]{bb99}
{Brotherton} M.~S.,  {van Breugel} W.,  {Stanford} S.~A.,  {Smith} R.~J.,
  {Boyle} B.~J.,  {Miller} L.,  {Shanks} T.,  {Croom} S.~M.,    {Filippenko}
  A.~V.,  1999, \apj, 520, L87

\bibitem[\protect\citeauthoryear{{Cannon} \& {Pickering}}{{Cannon} \&
  {Pickering}}{1924}]{hdcat}
{Cannon} A.~J.,  {Pickering} E.~C.,  1918--1924, Annals Harvard Obs., 91--99

\bibitem[\protect\citeauthoryear{{Cowley}, {Cowley}, {Jaschek} \&
  {Jaschek}}{{Cowley} et~al.}{1969}]{ccjj}
{Cowley} A.,  {Cowley} C.,  {Jaschek} M.,    {Jaschek} C.,  1969, \aj, 74, 375

\bibitem[\protect\citeauthoryear{{Crowther}}{{Crowther}}{1998}]{pc98}
{Crowther} P.~A.,  1998, in {Bedding} T.~R.,  {Booth} A.~J.,   {Davis} J.,
  eds, Fundamental Stellar Properties: The Interaction between Observation and
  Theory, IAU Symposium No. 189 Kluwer, Dordrecht, p.~137

\bibitem[\protect\citeauthoryear{{Dufton}, {McErlean}, {Lennon} \&
  {Ryans}}{{Dufton} et~al.}{2000}]{duf00}
{Dufton} P.~L.,  {McErlean} N.~D.,  {Lennon} D.~J.,    {Ryans} R. S.~I.,  2000,
  \aap, 353, 311

\bibitem[\protect\citeauthoryear{{Evans}, {Howarth} \& {Irwin}}{{Evans}
  et~al.}{2003}]{ehi}
{Evans} C.~J.,  {Howarth} I.~D.,    {Irwin} M.~J.,  2003, \mnras, submitted

\bibitem[\protect\citeauthoryear{{Fitzgerald}}{{Fitzgerald}}{1970}]{fitz70}
{Fitzgerald} M.~P.,  1970, \aap, 4, 234

\bibitem[\protect\citeauthoryear{{Gray} \& {Garrison}}{{Gray} \&
  {Garrison}}{1987}]{gg87}
{Gray} R.~O.,  {Garrison} R.~F.,  1987, \apjs, 65, 581

\bibitem[\protect\citeauthoryear{{Gray} \& {Garrison}}{{Gray} \&
  {Garrison}}{1989}]{gg89b}
{Gray} R.~O.,  {Garrison} R.~F.,  1989, \apjs, 70, 623

\bibitem[\protect\citeauthoryear{{Gray}, {Napier} \& {Winkler}}{{Gray}
  et~al.}{2001}]{gnw01}
{Gray} R.~O.,  {Napier} M.~G.,    {Winkler} L.~I.,  2001, \aj, 121, 2148

\bibitem[\protect\citeauthoryear{{Hartmann}}{{Hartmann}}{1904}]{har04}
{Hartmann} J.,  1904, \apj, 19, 268

\bibitem[\protect\citeauthoryear{{Haser}, {Pauldrach}, {Lennon}, {Kudritzki},
  {Lennon}, {Puls} \& {Voels}}{{Haser} et~al.}{1998}]{hpl98}
{Haser} S.~M.,  {Pauldrach} A. W.~A.,  {Lennon} D.~J.,  {Kudritzki} R.-P.,
  {Lennon} M.,  {Puls} J.,    {Voels} S.~A.,  1998, \aap, 330, 285

\bibitem[\protect\citeauthoryear{{Hendry}}{{Hendry}}{1981}]{h81}
{Hendry} E.~M.,  1981, \aj, 86, 1540

\bibitem[\protect\citeauthoryear{{Hill}}{{Hill}}{1999}]{hill99}
{Hill} V.,  1999, \aap, 345, 430

\bibitem[\protect\citeauthoryear{{Humphreys}}{{Humphreys}}{1978}]{hum78}
{Humphreys} R.~M.,  1978, \apjs, 38, 309

\bibitem[\protect\citeauthoryear{{Humphreys}}{{Humphreys}}{1983}]{hum83}
{Humphreys} R.~M.,  1983, \apj, 265, 176

\bibitem[\protect\citeauthoryear{{Humphreys}, {Kudritzki} \&
  {Groth}}{{Humphreys} et~al.}{1991}]{hkg91}
{Humphreys} R.~M.,  {Kudritzki} R.-P.,    {Groth} H.~G.,  1991, \aap, 245, 593

\bibitem[\protect\citeauthoryear{{Humphreys} \& {McElroy}}{{Humphreys} \&
  {McElroy}}{1984}]{hm84}
{Humphreys} R.~M.,  {McElroy} D.~B.,  1984, \apj, 284, 565

\bibitem[\protect\citeauthoryear{{Hutchings}}{{Hutchings}}{1966}]{h66}
{Hutchings} J.~B.,  1966, \mnras, 132, 433

\bibitem[\protect\citeauthoryear{{Jaschek} \& {Jaschek}}{{Jaschek} \&
  {Jaschek}}{1990}]{jj90}
{Jaschek} C.,  {Jaschek} M.,  1990, The Classification of Stars.
Cambridge University Press, Cambridge

\bibitem[\protect\citeauthoryear{{Kudritzki}, {Bresolin} \&
  {Przybilla}}{{Kudritzki} et~al.}{2003}]{k03}
{Kudritzki} R.-P.,  {Bresolin} F.,    {Przybilla} N.,  2003, \apj, 582, 83L

\bibitem[\protect\citeauthoryear{{Kurucz}}{{Kurucz}}{1991}]{k91}
{Kurucz} R.~L.,  1991, in {Crivellari} L.,  {Hubeny} I.,   {Hummer} D.~G.,
  eds, Stellar Atmospheres: Beyond Classical Models Kluwer, Dordrect, p.~441

\bibitem[\protect\citeauthoryear{{Lennon}}{{Lennon}}{1997}]{djl}
{Lennon} D.~J.,  1997, \aap, 317, 871

\bibitem[\protect\citeauthoryear{{Lewis}, {Cannon}, {Taylor}, {Glazebrook},
  {Waller}, {Whittard}, {Wilcox} \& {Willis}}{{Lewis} et~al.}{2002}]{2df}
{Lewis} I.~J.,  {Cannon} R.~D.,  {Taylor} K.,  {Glazebrook} K.,  {Waller}
  L.~G.,  {Whittard} J.~D.,  {Wilcox} J.~K.,    {Willis} K.~C.,  2002, \mnras,
  333, 279

\bibitem[\protect\citeauthoryear{{Luck}, {Moffett}, {Barnes} \&
  {Gieren}}{{Luck} et~al.}{1997}]{lmb97}
{Luck} R.~E.,  {Moffett} T.~J.,  {Barnes} T.~G.,    {Gieren} W.~P.,  1997, \aj,
  115, 605

\bibitem[\protect\citeauthoryear{{Maeder}, {Grebel} \& {Mermilliod}}{{Maeder}
  et~al.}{1999}]{mgm99}
{Maeder} A.,  {Grebel} E.~K.,    {Mermilliod} J.~C.,  1999, \aap, 346, 459

\bibitem[\protect\citeauthoryear{{Maury}}{{Maury}}{1897}]{maury97}
{Maury} A.~C.,  1897, Annals Harvard Obs., 28, 1

\bibitem[\protect\citeauthoryear{{McAlister}}{{McAlister}}{1979}]{mcali79}
{McAlister} H.~A.,  1979, \pasp, 90, 288

\bibitem[\protect\citeauthoryear{{McCarthy}, {Lennon}, {Venn}, {Kudritzki},
  {Puls} \& {Najarro}}{{McCarthy} et~al.}{1995}]{ml95}
{McCarthy} J.~K.,  {Lennon} D.~J.,  {Venn} K.~A.,  {Kudritzki} R.-P.,  {Puls}
  J.,    {Najarro} F.,  1995, \apj, 455, L135

\bibitem[\protect\citeauthoryear{{Mermilliod}, {Mermilliod} \&
  {Hauck}}{{Mermilliod} et~al.}{1997}]{mmh97}
{Mermilliod} J.-C.,  {Mermilliod} M.,    {Hauck} B.,  1997, \aaps, 124, 349

\bibitem[\protect\citeauthoryear{{Morgan} \& {Keenan}}{{Morgan} \&
  {Keenan}}{1973}]{mk73}
{Morgan} D.~H.,  {Keenan} P.~C.,  1973, \araap, 11, 29

\bibitem[\protect\citeauthoryear{{Morgan}}{{Morgan}}{1937}]{m37}
{Morgan} W.~W.,  1937, \apj, 85, 380

\bibitem[\protect\citeauthoryear{{Morgan}, {Code} \& {Whitford}}{{Morgan}
  et~al.}{1955}]{mcw55}
{Morgan} W.~W.,  {Code} A.~D.,    {Whitford} A.~E.,  1955, \apjs, 2, 41

\bibitem[\protect\citeauthoryear{{Morgan}, {Harris} \& {Johnson}}{{Morgan}
  et~al.}{1953}]{mhj53}
{Morgan} W.~W.,  {Harris} D.~L.,    {Johnson} H.~L.,  1953, \apj, 118, 92

\bibitem[\protect\citeauthoryear{{Morgan}, {Keenan} \& {Kellman}}{{Morgan}
  et~al.}{1943}]{mkk}
{Morgan} W.~W.,  {Keenan} P.~C.,    {Kellman} E.,  1943, An atlas of stellar
  spectra.
Chicago Univ. Press

\bibitem[\protect\citeauthoryear{{Morgan} \& {Roman}}{{Morgan} \&
  {Roman}}{1950}]{mr50}
{Morgan} W.~W.,  {Roman} N.~G.,  1950, \apj, 112, 362

\bibitem[\protect\citeauthoryear{{Pickering}}{{Pickering}}{1890}]{pick90b}
{Pickering} E.~C.,  1890, Annals Harvard Obs., 27, 1

\bibitem[\protect\citeauthoryear{{Pryzbilla}}{{Pryzbilla}}{2002}]{pryz02}
{Pryzbilla} N.,  2002, PhD thesis, Univ. of Munich

\bibitem[\protect\citeauthoryear{{Rolleston}, {Venn}, {Tolstoy} \&
  {Dufton}}{{Rolleston} et~al.}{2003}]{rvtd03}
{Rolleston} W. R.~J.,  {Venn} K.~A.,  {Tolstoy} E.,    {Dufton} P.~L.,  2003,
  \aap, 400, 21

\bibitem[\protect\citeauthoryear{{Russell} \& {Dopita}}{{Russell} \&
  {Dopita}}{1992}]{r92}
{Russell} S.~C.,  {Dopita} M.~A.,  1992, \apj, 384, 508

\bibitem[\protect\citeauthoryear{{Schmidt-Kaler}}{{Schmidt-Kaler}}{1982}]{sk82}
{Schmidt-Kaler} T.,  1982, in {Schaifers} K.,  {Voigt} H.~H.,  eds,
  Landolt-B$\ddot{o}$rnstein, Group VI, Vol 2b Springer-Verlag, p.~1

\bibitem[\protect\citeauthoryear{{Slettebak}}{{Slettebak}}{1954}]{slet54}
{Slettebak} A.,  1954, 1954, 119, 146

\bibitem[\protect\citeauthoryear{{Venn}}{{Venn}}{1995}]{venn95}
{Venn} K.~A.,  1995, \apjs, 99, 659

\bibitem[\protect\citeauthoryear{{Venn}}{{Venn}}{1999}]{venn99}
{Venn} K.~A.,  1999, \apj, 518, 405

\bibitem[\protect\citeauthoryear{{Venn}, {McCarthy}, {Lennon}, {Przybilla},
  {Kudritzki} \& {Lemke}}{{Venn} et~al.}{2000}]{venn00b}
{Venn} K.~A.,  {McCarthy} J.~K.,  {Lennon} D.~J.,  {Przybilla} N.,  {Kudritzki}
  R.-P.,    {Lemke} M.,  2000, \apj, 541, 610

\bibitem[\protect\citeauthoryear{{Verdugo}, {Talavera} \& {G$\acute{\rm o}$mez
  de Castro}}{{Verdugo} et~al.}{1999}]{vtg99b}
{Verdugo} E.,  {Talavera} A.,    {G$\acute{\rm o}$mez de Castro} A.~I.,  1999,
  \aap, 346, 819

\bibitem[\protect\citeauthoryear{{Walborn}}{{Walborn}}{1979}]{wal79}
{Walborn} N.~R.,  1979, in {McCarthy} M.~F.,  {Philip} A. G.~D.,   {Coyne}
  G.~V.,  eds, Spectral Classification of the Future, IAU Coll. 47 Vatican
  Observatory, p.~337

\bibitem[\protect\citeauthoryear{{Yamashita}, {Nariai} \&
  {Norimato}}{{Yamashita} et~al.}{1977}]{yam77}
{Yamashita} Y.,  {Nariai} K.,    {Norimato} Y.,  1977, An atlas of
  representative stellar spectra.
Univ. of Tokyo Press

\end{thebibliography}

\end{document}